\pdfoutput=1
\documentclass[superscriptaddress,showpacs,twocolumn,amsmath,amssymb,prl]{revtex4} 

\usepackage{graphicx} 
\usepackage{amsmath}
\usepackage{bm}
\usepackage{color}

\begin{document}


\title{A V-shape superconducting artificial atom based on two inductively coupled transmons\newline}


\author{\'E.~Dumur}
\affiliation
{Institut N\'eel, CNRS--Universit\'e Joseph Fourier, BP 166, 38042 Grenoble-cedex 9, France}

\author{B.~K\"ung}
\affiliation
{Institut N\'eel, CNRS--Universit\'e Joseph Fourier, BP 166, 38042 Grenoble-cedex 9, France}

\author{A.K.~Feofanov\footnote{Present address: \'Ecole Polytechnique F\'ed\'erale de Lausanne, 1015 Lausanne, Switzerland}}
\affiliation
{Institut N\'eel, CNRS--Universit\'e Joseph Fourier, BP 166, 38042 Grenoble-cedex 9, France}

\author{T.~Weissl}
\affiliation
{Institut N\'eel, CNRS--Universit\'e Joseph Fourier, BP 166, 38042 Grenoble-cedex 9, France}

\author{N.~Roch}
\affiliation
{Institut N\'eel, CNRS--Universit\'e Joseph Fourier, BP 166, 38042 Grenoble-cedex 9, France}

\author{C.~Naud}
\affiliation
{Institut N\'eel, CNRS--Universit\'e Joseph Fourier, BP 166, 38042 Grenoble-cedex 9, France}

\author{W.~Guichard}
\affiliation
{Institut N\'eel, CNRS--Universit\'e Joseph Fourier, BP 166, 38042 Grenoble-cedex 9, France}

\author{O.
Buisson}
\affiliation
{Institut N\'eel, CNRS--Universit\'e Joseph Fourier, BP 166, 38042 Grenoble-cedex 9, France}

\date{\today}


\begin{abstract}
Circuit quantum electrodynamics systems are typically built from resonators and two-level artificial atoms, but the use of multi-level artificial atoms instead can enable promising applications in quantum technology.
Here we present an implementation of a Josephson junction circuit dedicated to operate as a V-shape artificial atom.
Based on a concept of two internal degrees of freedom, the device consists of two transmon qubits coupled by an inductance.
The Josephson nonlinearity introduces a strong diagonal coupling between the two degrees of freedom that finds applications in quantum non-demolition readout schemes, and in the realization of microwave cross-Kerr media based on superconducting circuits.
\end{abstract}

\maketitle
Both in scientific and technological interest, the electromagnetic coupling of two-level systems and light has been the source of a great number of studies on quantum systems \cite{Haroche06}.
Replacing two-level by multi-level systems offers possibilities that go beyond the addition of complexity.
In experiments focusing on light and single photons, multi-level systems are at the origin of effects like electromagnetically induced transparency (EIT) \cite{Harris90} or the generation of entangled photon pairs \cite{Aspect81}.
In experiments focusing on two-level systems, advanced tools such as sideband cooling \cite{Monroe95} and state measurement \cite{Leibfried03, Jelezko04} become accessible when incorporating ancillary levels.

Inspired by these quantum experiments with natural atoms and ions, an adaptation to the field of superconducting circuits is an obvious line of research that has been followed using systems derived from qubits \cite{You11}.
In these systems with a single degree of freedom, selection rules are absent or favor ladder-shape level schemes.
Sideband cooling \cite{Valenzuela06} and EIT \cite{Abdumalikov10} have for instance been realized using flux qubits.
Related effects have also been studied in transmons and phase qubits \cite{Silanpaa09, Baur09}.

In comparison, only few theoretical studies have addressed the V-shape level scheme \cite{Gambetta11, Diniz13} despite its very successful application for quantum measurements in trapped ions \cite{Leibfried03} and nitrogen--vacancy centers in diamond \cite{Jelezko04}.
In this context, a V-shape system is understood as a qubit with good coherence properties formed by a ground and excited state $| g \rangle$ and $| e \rangle$, and an ancillary level $| a \rangle$ coupled to the ground state.
The transition between $| a \rangle$ and $| e \rangle$ should be suppressed.
Finally, transitions to higher states should be well out of resonance with the two principle transitions $|g\rangle \leftrightarrow |e\rangle$ and $|g\rangle \leftrightarrow |a\rangle$.
Combining these properties in a system with a single degree of freedom is challenging, but when using two degrees of freedom the separated $|g\rangle \leftrightarrow |e\rangle$ and $|g\rangle \leftrightarrow |a\rangle$ transitions naturally occur.
For quantum measurements, this offers the possibility to couple the readout transition $|g\rangle \leftrightarrow |a\rangle$ to the outside world, while keeping the qubit transition  $|g\rangle \leftrightarrow |e\rangle$ decoupled from it, and in particular protected from decay induced by the Purcell effect \cite{Gambetta11, Diniz13}.
Selective coupling is also a prerequisite for photon interaction schemes between spatially separated modes in which the multi-level device plays the role of a cross-Kerr medium \cite{Rebic09, Hu11, Neumeier13}.
V-shape properties were already mentionned in fluxonium~\cite{Manucharyan09}, phase qubit~\cite{Lecocq2012} and, tunable coupling qubit~\cite{Gambetta11, Srinivasan11, Hoffman2011} but, to our knowledge, not yet verified experimentally.
In this paper, we present a demonstration of a V-shape superconducting artificial atom verifying the ensemble of properties listed previously.

Our system is based on two inductively coupled transmons and the two degrees of freedom are given by two normal modes of the circuit \cite{Lecocq11squid}.
Their frequencies can be freely chosen by design.
The characteristics of the two modes predestine them to play the roles of a logical qubit and an ancillary qubit (or simply: qubit and ancilla).
The qubit part is played by the low-frequency mode which shows good coherence properties and a large anharmonicity, as it is equivalent to the well-established transmon.
In a circuit quantum electrodynamics architecture \cite{Wallraff04}, this mode couples to the photon field in a nearby resonator via its electric dipole moment.
The ancilla part is played by the high-frequency mode magnetically coupled to the resonator.

Interestingly the Josephson nonlinearity is at the origin of the V-shape property.
Usually the Josephson nonlinearity produces a strong anharmonicity in the low-frequency mode of a superconducting circuit.
This prevents from contamination by higher energy states and thus reduces the quantum dynamics to those of a two-level system \cite{Claudon08}.
In our device, the Josephson nonlinearity generates in addition a cross-anharmonicity effect between the two modes which induces a diagonal coupling \cite{Plantenberg07}.
This coupling leads to a frequency shift of one mode of more than $100 \, \mathrm{MHz}$ conditional on the excitation state of the other mode.
Thus this cross-anharmonicity prevents from contamination by the fourth state in which there is one excitation in each mode.
In that way the system dynamics are reduced to those of a V-shape system.

\begin{figure}[tb]
\centering
\includegraphics[width=\columnwidth]{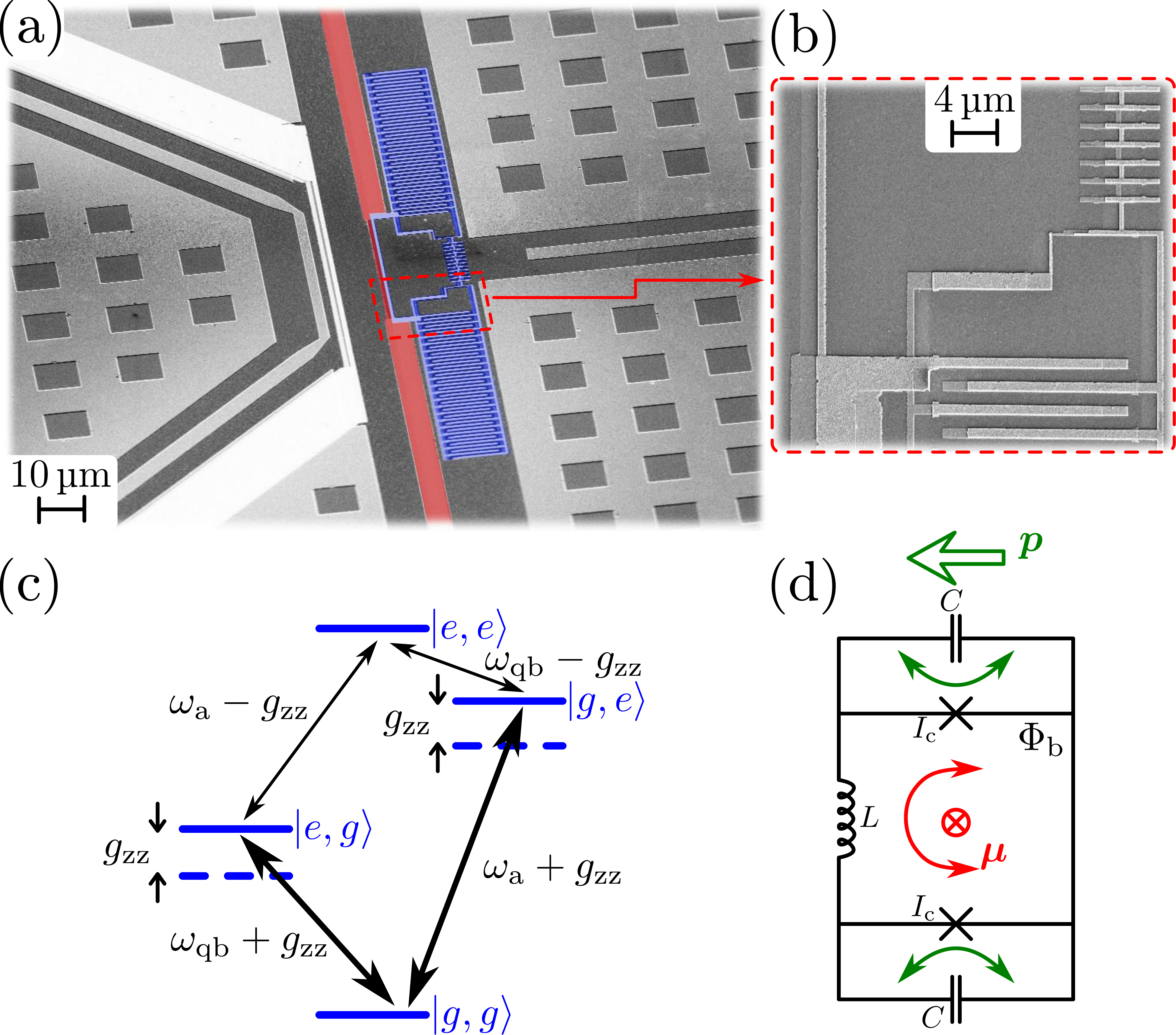}
\caption{(Color) (a) False-colored scanning electron micrograph of the sample.
The V-shape qubit circuit (blue) is coupled to the resonator (red) via a shared inductor.
 (b) Magnified view of the core part showing parts of the chain, a small junction, and parts of the capacitor (from top to bottom).
(c) Energy level diagram of the V-shape qubit.
Solid levels show the energy levels of the coupled perturbative Hamiltonian $\mathcal{H}$ in Eq.~\eqref{eq:VS_Hamiltonian_sigma}.
Dashed levels show the energy levels of the same Hamiltonian (up to an energy offset) without the coupling term $-\hbar g_{zz} \sigma^{\mathrm{(qb)}}_z \sigma^{\mathrm{(a)}}_z/2$.
Thanks to the energy shifts governed by the coupling strength $g_{zz}$, the lowest three levels can be addressed without populating the fourth level, and thus the experimental circuit realizes a V-shape level scheme (thick arrows) (d) Equivalent qubit circuit diagram consisting of two capacitances $C$, two Josephson junctions with critical current $I_c$, and an inductor $L$.
The current oscillations and the magnetic dipole moment $\bm{\mu}$ associated with the ancilla mode are indicated with red symbols, the current oscillations and the electric dipole moment $\bm{p}$ associated with the qubit mode are indicated with green symbols.
The loop is biased with a magnetic flux $\Phi_b$.}
\label{fig:VS_figure_sample}
\end{figure}

The transmon circuit that forms the basis of our circuit consists of a small Josephson junction with critical current $I_c$ that is shunted by an interdigital capacitance $C$.
We introduce the associated Josephson and charging energies as $E_J = \Phi_0 I_c/(2\pi)$ and $E_C = (2e)^2/(2C)$, where $\Phi_0 = h/(2e)$ is the magnetic flux quantum.
We couple two identical transmons by integrating their Josephson junctions into a loop with a large linear inductance $L$.
The magnitude of $L$ is comparable to the Josephson inductance $L_J = \Phi_0/(2 \pi I_c)$.
The device shown in Fig.~\ref{fig:VS_figure_sample}(a,b) is fabricated from thin-film aluminium on a high-resistivity silicon substrate.
After patterning the larger parts of the structure by electron beam lithography and wet etching, the qubit structure as well as the center conductor of a coplanar-waveguide resonator are fabricated by lift-off using the controlled-undercut technique \cite{Lecocq11ebeam}.
The linear inductance is realized in the form of a chain of twelve large Josephson junctions \cite{Manucharyan09} of critical current $I_c' \gg I_c$ such that $L = 12 \times \Phi_0/(2 \pi I_c')$.

The simplified diagram of our circuit is shown in Fig.~\ref{fig:VS_figure_sample}(d).
The currents $I_1$ and $I_2$ through the two small junctions represent the two degrees of freedom in the circuit.
The circuit exhibits two modes: a symmetric one corresponding to an in-phase oscillation of $I_1$ and $I_2$, and an antisymmetric one corresponding to an oscillation of $I_1$ and $I_2$ in anti-phase.
The symmetric (or qubit) mode corresponds to the plasma oscillation of the superconducting quantum interference device (SQUID) formed by the two junctions in a superconducting loop.
Its electric dipole moment points in line of the junctions, as indicated in the diagram, whereas its magnetic dipole moment is zero.
The antisymmetric (or ancilla) mode is usually not accessible in a SQUID due to its elevated frequency.
In our device however, the large inductance $L$ ensures that the frequency of this mode falls within the measurement bandwidth.
Its magnetic dipole moment points out of the circuit plane, whereas its electric dipole moment is zero.
These two orthogonal dipoles enable selective coupling between the qubit and the ancilla, opening the way to novel circuit architecture possibilities \cite{Diniz13,Hu11,Neumeier13}.




The SQUID flux bias $\Phi_b$ controls the mode energies and their mutual coupling.
An optimal point for the operation as a V-shape device is reached at the ``sweet spot'' $\Phi_b = 0$, which we assume for the following theoretical description \cite{Lecocq11squid}.
Both modes are anharmonic and we may consider them as two-level systems with transition energies $\hbar \omega_\mathrm{qb} \approx \sqrt{2 E_J E_C}$ and $\hbar \omega_\mathrm{a} \approx \sqrt{2 E_J E_C} \sqrt{1+2L_J/L}$.
Remarkably, the two modes are coupled by a $\sigma_z\sigma_z$ term, whereas other coupling terms are absent at zero flux due to symmetry reasons.
This follows from the perturbative treatment of the full Hamiltonian of the circuit depicted in Fig.~\ref{fig:VS_figure_sample}(d).
For the purpose of this paper, we will thus describe the system by the simplified Hamiltonian
\begin{equation}
\label{eq:VS_Hamiltonian_sigma}
\mathcal{H} =
  \hbar \omega_\mathrm{qb} \sigma^{\mathrm{(qb)}}_z/2
+ \hbar \omega_\mathrm{a}  \sigma^{\mathrm{(a)}}_z/2
- \hbar g_{zz} \sigma^{\mathrm{(qb)}}_z \sigma^{\mathrm{(a)}}_z/2,
\end{equation}
where $\sigma^{\mathrm{(qb)}}_z$ ($\sigma^{\mathrm{(a)}}_z$) are Pauli matrices of the qubit (ancilla).
The cross-anharmonicity is expressed as
\begin{equation}
\label{eq:VS_gzz}
\hbar g_{zz} = \frac{E_C}{8\sqrt{1 + 2L_J/L}}.
\end{equation}

In Fig.~\ref{fig:VS_figure_sample}(c), we show an energy level diagram of our system to clarify the role of the coupling term.
The eigenstates of the uncoupled Hamiltonian $\hbar \omega_\mathrm{qb} \sigma^{\mathrm{(qb)}}_z/2 + \hbar \omega_\mathrm{a}  \sigma^{\mathrm{(a)}}_z/2$ are shown as dashed levels.
The solid levels show the eigenenergies $E_{i,j}$ of the coupled Hamiltonian given in Eq.~\eqref{eq:VS_Hamiltonian_sigma} with eigenstates $|i,j\rangle$, where $i \, (j) = g, \, e$ denotes the qubit (ancilla) state.
The transition energy $E_{e,g} - E_{g,g}$ is detuned from $E_{e,e} - E_{g,e}$ by an amount $2\hbar g_{zz}$.
Equally, the transition energy $E_{g,e} - E_{g,g}$ is detuned from $E_{e,e} - E_{e,g}$ by $2\hbar g_{zz}$.
As long as $\hbar g_{zz}$ is large enough, it will allow for V-shape system dynamics in which only the states $|g,g \rangle = | g \rangle$, $|e,g \rangle = | e \rangle$, and $|g,e \rangle = | a \rangle$ play a role.
In particular, $ g_{zz}$ must be significantly larger than the linewidths of the qubit and ancilla transitions.

In the following, we describe the measurements performed to determine the mode energies and to demonstrate the cross-anharmonicity in our device.
Our sample is placed in a dilution refrigerator at a temperature of approximately $30 \, \mathrm{mK}$.
The quantum circuit is coupled to a coplanar-waveguide resonator through an inductance shared by the qubit loop and the resonator \cite{Abdumalikov10} as well as through stray capacitances.
By placing the circuit at the grounded end of a quarter-wave resonator, we can achieve the interesting configuration in which the inductive coupling between ancilla and resonator is maximized whereas the capacitive coupling between qubit and resonator is eliminated at all frequencies.
This allows for a fast qubit readout protocol free of decay induced by the Purcell effect \cite{Diniz13}.
In our device, this configuration is realized within geometrical constraints, leading to a coupling between qubit and resonator small enough to suppress Purcell decay but large enough to allow exciting the qubit with a microwave tone sent through the resonator.

At its open end, the resonator is capacitively coupled to a microwave transmission line through which we measure the transmission of a readout tone at frequency $f_{ro}$ close to the resonator frequency $\omega_r/2\pi \approx 7.2 \, \mathrm{GHz}$.
Through the presence of a SQUID in the center conductor \cite{Osborn07}, $\omega_r$ is tunable by magnetic field over a range of $\sim 150 \, \mathrm{MHz}$.
The input signal is attenuated by $20 \, \mathrm{dB}$ at $4.2 \, \mathrm{K}$ and by $40 \, \mathrm{dB}$ at base temperature and the output signal passes through a low-noise amplifier at $4.2 \, \mathrm{K}$.
The sample is protected from amplifier noise by two circulators and a low-pass filter.
A second sample consisting of a resonator at $7.7 \, \mathrm{GHz}$ and a V-shape device are present on the same chip.
Thanks to the well-separated resonator frequencies, we can independently measure the two samples using the same transmission line.

\begin{figure}[tb]
\centering
\includegraphics[width=0.9\columnwidth]{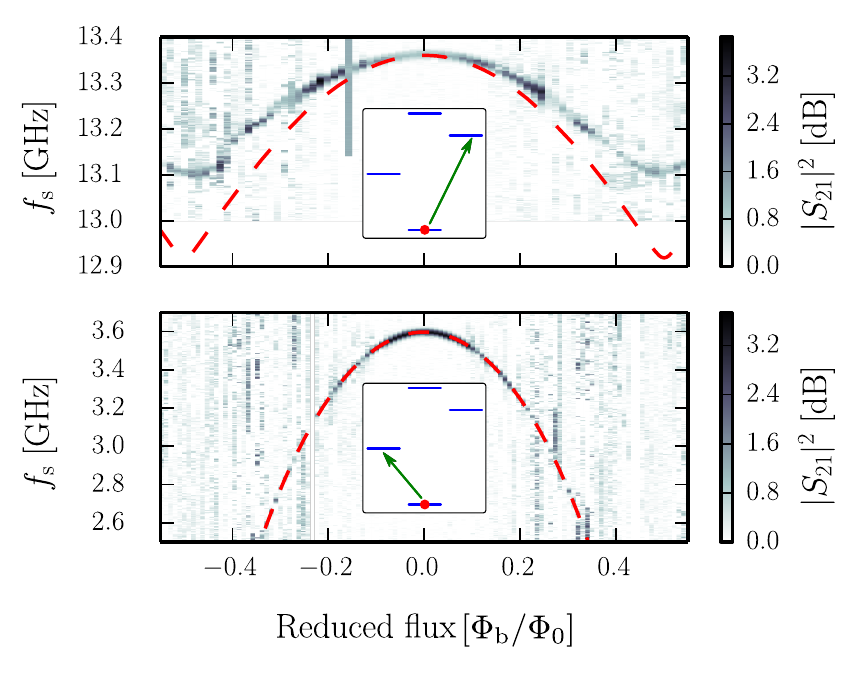}
\caption{(Color) Spectroscopy of the V-shape artificial atom.
The gray scale encodes the transmission of a readout tone close to the resonator frequency $\omega_r/(2\pi)$ in the presence of a spectroscopy tone whose frequency $f_s$ is swept.
The magnetic field is varied along the horizontal axis and converted to flux $\Phi_b$ through the SQUID loop.
The gray lines represent the transition of the ancilla in the top graph and the transition of the qubit in the bottom graph as illustrated in the insets.
Dashed lines show numerical model calculations of these transition energies.
The small discrepancy on the ancilla spectroscopy between experiment and prediction close to $\Phi_b/\Phi_0 \approx \pm 1/2$ may be explained by taking into account a $35$~\% asymetrical critical current in the two coupled transmons.}
\label{fig:VS_figure_spectroscopy}
\end{figure}

We performed two-tone spectroscopy to map out the energy diagram of the artificial atom shown in Fig.~\ref{fig:VS_figure_spectroscopy}.
We send a spectroscopy tone at frequency $f_s$ through the transmission line while measuring transmission at $f_{ro}$.
Both the qubit and the ancilla are dispersively coupled to the resonator.
The excitation of their transitions leads to a shift in $\omega_r$ of a few MHz, and consequently to a change in transmission at $f_{ro}$ \cite{Wallraff04}.

As mentioned, the dedicated operation point is at zero flux, but for the purpose of characterization we study the full flux dependence.
We vary the external magnetic field with a small coil around the sample.
In the data in Fig.~\ref{fig:VS_figure_spectroscopy} we distinguish two dark lines corresponding to the qubit transition $\hbar\omega_\mathrm{qb}  + \hbar g_{zz} = E_{e,g} - E_{g,g}$ (bottom) and the ancilla transition  $\hbar\omega_\mathrm{a}  + \hbar g_{zz} = E_{g,e} - E_{g,g}$, respectively.
As a function of flux, the frequency $\omega_\mathrm{qb}$ of the qubit mode varies more strongly on a relative scale than that of the ancilla mode, $\omega_\mathrm{a}$.
The ancilla mode involves principally the elements $L$ and $C$ that are insensitive to flux.
In contrast, the qubit mode does not involve $L$ and its frequency is expected to drop strongly as the flux bias reaches $\Phi_b = \Phi_0/2$.

In order to compare the circuit model in Fig.~\ref{fig:VS_figure_sample}(d) with experiment, we performed numerical calculations of the spectrum of its full Hamiltonian derived in Ref.~\cite{Lecocq11squid}.
The model depends on the three circuit parameters $I_c$, $C$, and $L$ which we take as fitting parameters.
The numerical calculation consists of a solution of the discretized Schr\"odinger equation using the Kwant code \cite{Groth14}.
The results of these calculations are shown as dashed lines in Fig.~\ref{fig:VS_figure_spectroscopy}.
The fit to the experimental data yields the parameters $I_c = 8.19 \, \mathrm{nA}$, $C = 39.7 \, \mathrm{fF}$, and $L = 0.192 \times L_J$.

Due to mode anharmonicity, the transition from the first to the second excited state of the qubit is detuned from $\omega_\mathrm{qb}$ by $-0.30 \, \mathrm{GHz}$, as measured via two-photon excitation into that state at large microwave power.
We performed measurements of Rabi oscillations and qubit relaxation to estimate a coherence time of $T_{2,\mathrm{Rabi}} = 0.5 \, \mathrm{\mu s}$ and a lifetime $T_1 = 0.6 \, \mathrm{\mu s}$ of the qubit.

\begin{figure}[tb]
\centering
\includegraphics[width=\columnwidth]{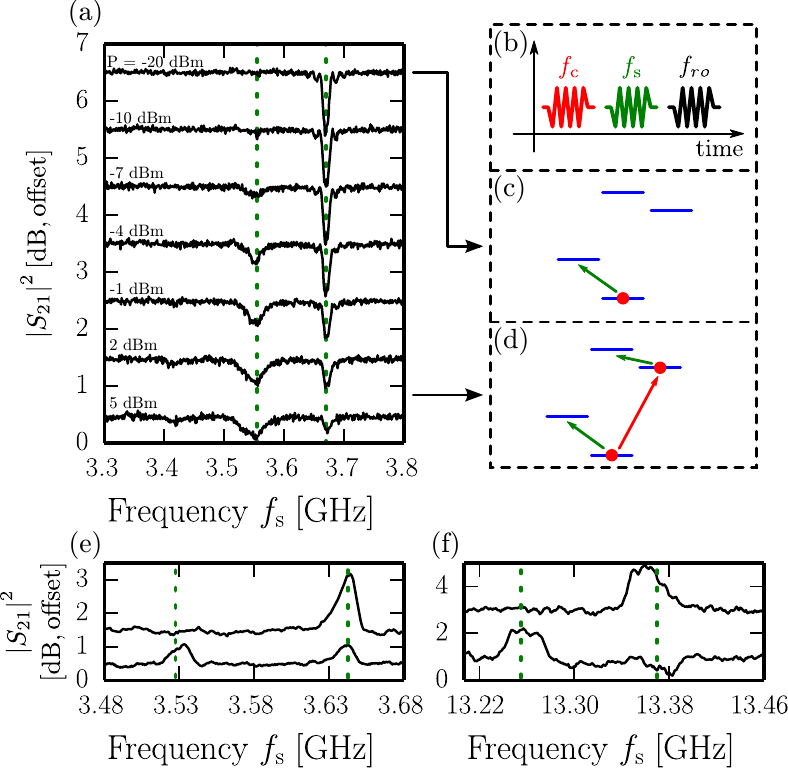}
\caption{(Color) (a) Measurement of the cross-anharmonicity $g_{zz}$ based on a sequence involving a conditioning pulse at $f_c = \omega_\mathrm{a}/(2\pi)$ to populate the ancilla, a spectroscopy pulse at $f_s$ to scan over the qubit transition, and a readout pulse at $f_{ro}$, cf.~the scheme in (b).
From the topmost to the lowest curve, we increase the power of the conditioning pulse in steps.
Below the qubit dip present for all powers at $\omega_\mathrm{qb}/(2\pi) \approx 3.67 \, \mathrm{GHz}$, a second dip emerges for high powers.
The peak separation is a measure of the cross-anharmonicity.
The situations without and with the first pulse are represented in the diagrams in (c) and (d).
(e) Control measurement of the peak separation on the twin sample on the same chip.
This measurement is performed with continuous tones, with a tone at $f_c = \omega_\mathrm{a}/(2\pi)$ turned off for the top trace, and turned on for the bottom trace.
The result is consistent with that shown in (a).
(f) Second control measurement with inverted roles of qubit and ancilla, i.e., the spectroscopy tone frequency is swept around the ancilla frequency, whereas the conditioning tone is resonant with the qubit at $f_c = \omega_\mathrm{qb}/(2\pi)$.
}
\label{fig:VS_figure_gzz}
\end{figure}

In order to observe the cross-anharmonicity $g_{zz}$ at zero flux, we choose the experimental approach of detecting changes in the qubit transition frequency depending on the presence or absence of a microwave excitation of the ancilla.
In essence, we perform a spectroscopy, in which we apply a spectroscopy pulse (frequency  $f_s$, duration $80 \, \mathrm{ns}$), followed by a readout pulse (frequency  $f_{ro}$, duration $250 \, \mathrm{ns}$).
Before these two pulses, we apply a conditioning pulse of duration $10 \, \mathrm{ns}$ and frequency $f_c$ resonant with the ancilla transition $|g,g \rangle \rightarrow |g,e \rangle$.
In the measurement shown in Fig.~\ref{fig:VS_figure_gzz}(a), the power of the conditioning pulse is gradually increased from top to bottom.
The pulse sequence is sketched in Fig.~\ref{fig:VS_figure_gzz}(b).
In the top trace in Fig.~\ref{fig:VS_figure_gzz}(a), we observe a single dip corresponding to the qubit transition $|g,g \rangle \rightarrow |e,g \rangle$ indicated in the level scheme in panel (c).
Towards the bottom, a second dip emerges next to the first one.
As indicated in the level scheme in panel (d), the second dip is caused by the population of the excited ancilla state $|g,e\rangle$ by the conditioning pulse (red arrow).
The transition $|g,e \rangle \rightarrow |e,e \rangle$ thus becomes available.
The depth of the two dips is ideally proportional to the occupation of the states $|g,g\rangle$ and $|g,e\rangle$, their frequencies are separated by $g_{zz}/\pi$.
This picture corresponds well to the observation, and we can extract a cross-anharmonicity of $g_{zz}/\pi = 115 \, \mathrm{MHz}$ (dashed lines).

We performed two further measurement to test the reproducibility and consistency of this result.
Firstly, we performed a measurement on the other, nominally identical V-shape device on the same chip, secondly, we interchanged the role of the qubit and the ancilla.
We could obtain a better overall data quality when measuring in continuous rather than pulsed mode, meaning that a measurement tone, a spectroscopy tone, and a conditioning tone were applied simultaneously.
In Fig.~\ref{fig:VS_figure_gzz}(e), we plot the first measurement showing a spectroscopy around the qubit frequency, with the conditioning tone at the ancilla frequency turned off (top trace), or turned on (bottom trace).
The result is largely consistent with that in Fig.~\ref{fig:VS_figure_gzz}(a), with a slightly smaller peak separation of about $110 \, \mathrm{MHz}$.
In Fig.~\ref{fig:VS_figure_gzz}(f), we plot a measurement with inverted roles of qubit and ancilla.
We sweep the spectroscopy frequency $f_s$ around the ancilla frequency, and excite the qubit with the conditioning tone at $\omega_\mathrm{qb}/(2\pi)$.
We observe a quantitatively similar behavior to the previous measurements with a peak separation of about $110 \, \mathrm{MHz}$.
The experimental value of $g_{zz}/\pi$ compares well to the theoretical value $144 \, \mathrm{MHz}$ obtained from Eq.~\eqref{eq:VS_gzz}, corroborating the validity of our circuit model.

With this result we demonstrate the nature and strength of the cross-anharmonicity in our device.
On this basis we can qualify our circuit as a V-shape system.
The measurement of the cross-anharmonicity and its interpretation are based on a solid theoretical model verified by spectroscopy.
Our circuit furthermore exhibits the good coherence properties required for the use in quantum experiments.
Integration in existing setups is straightforward thanks to a device technology proven in a wide range of circuit quantum electrodynamics experiments.
Among the applications, we highlight fast qubit readout techniques \cite{Diniz13} and cross-Kerr interactions at the few-photon level \cite{Hu11}.
Thanks to its electrical and magnetic dipole allowing a selective coupling of the ancilla and logical qubit, our circuit also fulfilled the requirements of the single-photon transistor proposed by Ref. \cite{Neumeier13}.

The authors thank  M.~Hofheinz, B.~Huard, Y.~Kubo, I.~Matei, and A.~Wallraff for fruitful discussions.
The research has been supported by European IP SOLID and ANR-NSFC QUEXSUPERC.
B.K and T.W.
acknowledge support from the Swiss NSF and Grenoble Nanoscience Foundation, respectively.
W.G.
is supported by Institut Universitaire de France and by the European Research Council (grant no.
306731).



\begin{thebibliography}{25}
\expandafter\ifx\csname natexlab\endcsname\relax\def\natexlab#1{#1}\fi
\expandafter\ifx\csname bibnamefont\endcsname\relax
  \def\bibnamefont#1{#1}\fi
\expandafter\ifx\csname bibfnamefont\endcsname\relax
  \def\bibfnamefont#1{#1}\fi
\expandafter\ifx\csname citenamefont\endcsname\relax
  \def\citenamefont#1{#1}\fi
\expandafter\ifx\csname url\endcsname\relax
  \def\url#1{\texttt{#1}}\fi
\expandafter\ifx\csname urlprefix\endcsname\relax\def\urlprefix{URL }\fi
\providecommand{\bibinfo}[2]{#2}
\providecommand{\eprint}[2][]{\url{#2}}

\bibitem[{\citenamefont{Haroche and Raimond}(2006)}]{Haroche06}
\bibinfo{author}{\bibfnamefont{S.}~\bibnamefont{Haroche}} \bibnamefont{and}
  \bibinfo{author}{\bibfnamefont{J.-M.} \bibnamefont{Raimond}},
  \emph{\bibinfo{title}{Exploring the Quantum: Atoms, Cavities, and Photons}}
  (\bibinfo{publisher}{Oxford University Press}, \bibinfo{address}{Oxford},
  \bibinfo{year}{2006}).

\bibitem[{\citenamefont{Harris et~al.}(1990)\citenamefont{Harris, Field, and
  Imamo\ifmmode~\breve{g}\else \u{g}\fi{}lu}}]{Harris90}
\bibinfo{author}{\bibfnamefont{S.~E.} \bibnamefont{Harris}},
  \bibinfo{author}{\bibfnamefont{J.~E.} \bibnamefont{Field}}, \bibnamefont{and}
  \bibinfo{author}{\bibfnamefont{A.}~\bibnamefont{Imamo\ifmmode~\breve{g}\else
  \u{g}\fi{}lu}}, \bibinfo{journal}{Phys. Rev. Lett.}
  \textbf{\bibinfo{volume}{64}}, \bibinfo{pages}{1107} (\bibinfo{year}{1990}).

\bibitem[{\citenamefont{Aspect et~al.}(1981)\citenamefont{Aspect, Grangier, and
  Roger}}]{Aspect81}
\bibinfo{author}{\bibfnamefont{A.}~\bibnamefont{Aspect}},
  \bibinfo{author}{\bibfnamefont{P.}~\bibnamefont{Grangier}}, \bibnamefont{and}
  \bibinfo{author}{\bibfnamefont{G.}~\bibnamefont{Roger}},
  \bibinfo{journal}{Phys. Rev. Lett.} \textbf{\bibinfo{volume}{47}},
  \bibinfo{pages}{460} (\bibinfo{year}{1981}).

\bibitem[{\citenamefont{Monroe et~al.}(1995)\citenamefont{Monroe, Meekhof,
  King, Jefferts, Itano, Wineland, and Gould}}]{Monroe95}
\bibinfo{author}{\bibfnamefont{C.}~\bibnamefont{Monroe}},
  \bibinfo{author}{\bibfnamefont{D.~M.} \bibnamefont{Meekhof}},
  \bibinfo{author}{\bibfnamefont{B.~E.} \bibnamefont{King}},
  \bibinfo{author}{\bibfnamefont{S.~R.} \bibnamefont{Jefferts}},
  \bibinfo{author}{\bibfnamefont{W.~M.} \bibnamefont{Itano}},
  \bibinfo{author}{\bibfnamefont{D.~J.} \bibnamefont{Wineland}},
  \bibnamefont{and} \bibinfo{author}{\bibfnamefont{P.}~\bibnamefont{Gould}},
  \bibinfo{journal}{Phys. Rev. Lett.} \textbf{\bibinfo{volume}{75}},
  \bibinfo{pages}{4011} (\bibinfo{year}{1995}).

\bibitem[{\citenamefont{Leibfried et~al.}(2003)\citenamefont{Leibfried, Blatt,
  Monroe, and Wineland}}]{Leibfried03}
\bibinfo{author}{\bibfnamefont{D.}~\bibnamefont{Leibfried}},
  \bibinfo{author}{\bibfnamefont{R.}~\bibnamefont{Blatt}},
  \bibinfo{author}{\bibfnamefont{C.}~\bibnamefont{Monroe}}, \bibnamefont{and}
  \bibinfo{author}{\bibfnamefont{D.}~\bibnamefont{Wineland}},
  \bibinfo{journal}{Rev.
Mod.
Phys.} \textbf{\bibinfo{volume}{75}},
  \bibinfo{pages}{281} (\bibinfo{year}{2003}).

\bibitem[{\citenamefont{Jelezko et~al.}(2004)\citenamefont{Jelezko, Gaebel,
  Popa, Gruber, and Wrachtrup}}]{Jelezko04}
\bibinfo{author}{\bibfnamefont{F.}~\bibnamefont{Jelezko}},
  \bibinfo{author}{\bibfnamefont{T.}~\bibnamefont{Gaebel}},
  \bibinfo{author}{\bibfnamefont{I.}~\bibnamefont{Popa}},
  \bibinfo{author}{\bibfnamefont{A.}~\bibnamefont{Gruber}}, \bibnamefont{and}
  \bibinfo{author}{\bibfnamefont{J.}~\bibnamefont{Wrachtrup}},
  \bibinfo{journal}{Phys. Rev. Lett.} \textbf{\bibinfo{volume}{92}},
  \bibinfo{pages}{076401} (\bibinfo{year}{2004}).

\bibitem[{\citenamefont{You and Nori}(2011)}]{You11}
\bibinfo{author}{\bibfnamefont{J.~Q.} \bibnamefont{You}} \bibnamefont{and}
  \bibinfo{author}{\bibfnamefont{F.}~\bibnamefont{Nori}},
  \bibinfo{journal}{Nature (London)} \textbf{\bibinfo{volume}{474}},
  \bibinfo{pages}{589} (\bibinfo{year}{2011}).

\bibitem[{\citenamefont{Valenzuela et~al.}(2006)\citenamefont{Valenzuela,
  Oliver, Berns, Berggren, Levitov, and Orlando}}]{Valenzuela06}
\bibinfo{author}{\bibfnamefont{S.~O.} \bibnamefont{Valenzuela}},
  \bibinfo{author}{\bibfnamefont{W.~D.} \bibnamefont{Oliver}},
  \bibinfo{author}{\bibfnamefont{D.~M.} \bibnamefont{Berns}},
  \bibinfo{author}{\bibfnamefont{K.~K.} \bibnamefont{Berggren}},
  \bibinfo{author}{\bibfnamefont{L.~S.} \bibnamefont{Levitov}},
  \bibnamefont{and} \bibinfo{author}{\bibfnamefont{T.~P.}
  \bibnamefont{Orlando}}, \bibinfo{journal}{Science}
  \textbf{\bibinfo{volume}{314}}, \bibinfo{pages}{1589} (\bibinfo{year}{2006}).

\bibitem[{\citenamefont{Abdumalikov et~al.}(2010)\citenamefont{Abdumalikov,
  Astafiev, Zagoskin, Pashkin, Nakamura, and Tsai}}]{Abdumalikov10}
\bibinfo{author}{\bibfnamefont{A.~A.} \bibnamefont{Abdumalikov}},
  \bibinfo{author}{\bibfnamefont{O.}~\bibnamefont{Astafiev}},
  \bibinfo{author}{\bibfnamefont{A.~M.} \bibnamefont{Zagoskin}},
  \bibinfo{author}{\bibfnamefont{Y.~A.} \bibnamefont{Pashkin}},
  \bibinfo{author}{\bibfnamefont{Y.}~\bibnamefont{Nakamura}}, \bibnamefont{and}
  \bibinfo{author}{\bibfnamefont{J.~S.} \bibnamefont{Tsai}},
  \bibinfo{journal}{Phys. Rev. Lett.} \textbf{\bibinfo{volume}{104}},
  \bibinfo{pages}{193601} (\bibinfo{year}{2010}).

\bibitem[{\citenamefont{Sillanp\"a\"a et~al.}(2009)\citenamefont{Sillanp\"a\"a,
  Li, Cicak, Altomare, Park, Simmonds, Paraoanu, and Hakonen}}]{Silanpaa09}
\bibinfo{author}{\bibfnamefont{M.~A.} \bibnamefont{Sillanp\"a\"a}},
  \bibinfo{author}{\bibfnamefont{J.}~\bibnamefont{Li}},
  \bibinfo{author}{\bibfnamefont{K.}~\bibnamefont{Cicak}},
  \bibinfo{author}{\bibfnamefont{F.}~\bibnamefont{Altomare}},
  \bibinfo{author}{\bibfnamefont{J.~I.} \bibnamefont{Park}},
  \bibinfo{author}{\bibfnamefont{R.~W.} \bibnamefont{Simmonds}},
  \bibinfo{author}{\bibfnamefont{G.~S.} \bibnamefont{Paraoanu}},
  \bibnamefont{and} \bibinfo{author}{\bibfnamefont{P.~J.}
  \bibnamefont{Hakonen}}, \bibinfo{journal}{Phys. Rev. Lett.}
  \textbf{\bibinfo{volume}{103}}, \bibinfo{pages}{193601}
  (\bibinfo{year}{2009}).

\bibitem[{\citenamefont{Baur et~al.}(2009)\citenamefont{Baur, Filipp,
  Bianchetti, Fink, G\"oppl, Steffen, Leek, Blais, and Wallraff}}]{Baur09}
\bibinfo{author}{\bibfnamefont{M.}~\bibnamefont{Baur}},
  \bibinfo{author}{\bibfnamefont{S.}~\bibnamefont{Filipp}},
  \bibinfo{author}{\bibfnamefont{R.}~\bibnamefont{Bianchetti}},
  \bibinfo{author}{\bibfnamefont{J.~M.} \bibnamefont{Fink}},
  \bibinfo{author}{\bibfnamefont{M.}~\bibnamefont{G\"oppl}},
  \bibinfo{author}{\bibfnamefont{L.}~\bibnamefont{Steffen}},
  \bibinfo{author}{\bibfnamefont{P.~J.} \bibnamefont{Leek}},
  \bibinfo{author}{\bibfnamefont{A.}~\bibnamefont{Blais}}, \bibnamefont{and}
  \bibinfo{author}{\bibfnamefont{A.}~\bibnamefont{Wallraff}},
  \bibinfo{journal}{Phys. Rev. Lett.} \textbf{\bibinfo{volume}{102}},
  \bibinfo{pages}{243602} (\bibinfo{year}{2009}).

\bibitem[{\citenamefont{Gambetta et~al.}(2011)\citenamefont{Gambetta, Houck,
  and Blais}}]{Gambetta11}
\bibinfo{author}{\bibfnamefont{J.~M.} \bibnamefont{Gambetta}},
  \bibinfo{author}{\bibfnamefont{A.~A.} \bibnamefont{Houck}}, \bibnamefont{and}
  \bibinfo{author}{\bibfnamefont{A.}~\bibnamefont{Blais}},
  \bibinfo{journal}{Phys. Rev. Lett.} \textbf{\bibinfo{volume}{106}},
  \bibinfo{pages}{030502} (\bibinfo{year}{2011}).

\bibitem[{\citenamefont{Diniz et~al.}(2013)\citenamefont{Diniz, Dumur, Buisson,
  and Auff\`eves}}]{Diniz13}
\bibinfo{author}{\bibfnamefont{I.}~\bibnamefont{Diniz}},
  \bibinfo{author}{\bibfnamefont{\'E.}~\bibnamefont{Dumur}},
  \bibinfo{author}{\bibfnamefont{O.}~\bibnamefont{Buisson}}, \bibnamefont{and}
  \bibinfo{author}{\bibfnamefont{A.}~\bibnamefont{Auff\`eves}},
  \bibinfo{journal}{Phys. Rev. A} \textbf{\bibinfo{volume}{87}},
  \bibinfo{pages}{033837} (\bibinfo{year}{2013}).

\bibitem[{\citenamefont{Rebi\'{c} et~al.}(2009)\citenamefont{Rebi\'{c},
  Twamley, and Milburn}}]{Rebic09}
\bibinfo{author}{\bibfnamefont{S.}~\bibnamefont{Rebi\'{c}}},
  \bibinfo{author}{\bibfnamefont{J.}~\bibnamefont{Twamley}}, \bibnamefont{and}
  \bibinfo{author}{\bibfnamefont{G.~J.} \bibnamefont{Milburn}},
  \bibinfo{journal}{Phys. Rev. Lett.} \textbf{\bibinfo{volume}{103}},
  \bibinfo{pages}{150503} (\bibinfo{year}{2009}).

\bibitem[{\citenamefont{Hu et~al.}(2011)\citenamefont{Hu, Ge, Chen, Yang, and
  Chen}}]{Hu11}
\bibinfo{author}{\bibfnamefont{Y.}~\bibnamefont{Hu}},
  \bibinfo{author}{\bibfnamefont{G.-Q.} \bibnamefont{Ge}},
  \bibinfo{author}{\bibfnamefont{S.}~\bibnamefont{Chen}},
  \bibinfo{author}{\bibfnamefont{X.-F.} \bibnamefont{Yang}}, \bibnamefont{and}
  \bibinfo{author}{\bibfnamefont{Y.-L.} \bibnamefont{Chen}},
  \bibinfo{journal}{Phys. Rev. A} \textbf{\bibinfo{volume}{84}},
  \bibinfo{pages}{012329} (\bibinfo{year}{2011}).

\bibitem[{\citenamefont{Neumeier et~al.}(2013)\citenamefont{Neumeier, Leib, and
  Hartmann}}]{Neumeier13}
\bibinfo{author}{\bibfnamefont{L.}~\bibnamefont{Neumeier}},
  \bibinfo{author}{\bibfnamefont{M.}~\bibnamefont{Leib}}, \bibnamefont{and}
  \bibinfo{author}{\bibfnamefont{M.~J.} \bibnamefont{Hartmann}},
  \bibinfo{journal}{Phys. Rev. Lett.} \textbf{\bibinfo{volume}{111}},
  \bibinfo{pages}{063601} (\bibinfo{year}{2013}).

\bibitem[{\citenamefont{Manucharyan et~al.}(2009)\citenamefont{Manucharyan,
  Koch, Glazman, and Devoret}}]{Manucharyan09}
\bibinfo{author}{\bibfnamefont{V.~E.} \bibnamefont{Manucharyan}},
  \bibinfo{author}{\bibfnamefont{J.}~\bibnamefont{Koch}},
  \bibinfo{author}{\bibfnamefont{L.~I.} \bibnamefont{Glazman}},
  \bibnamefont{and} \bibinfo{author}{\bibfnamefont{M.~H.}
  \bibnamefont{Devoret}}, \bibinfo{journal}{Science}
  \textbf{\bibinfo{volume}{326}}, \bibinfo{pages}{113} (\bibinfo{year}{2009}).

\bibitem[{\citenamefont{Lecocq et~al.}(2011)\citenamefont{Lecocq, Pop, Matei,
Dumur, Feofanov, Naud, Guichard, and Buisson}}]{Lecocq2012}
\bibinfo{author}{\bibfnamefont{F.}~\bibnamefont{Lecocq}},
  \bibinfo{author}{\bibfnamefont{I.~M.}~\bibnamefont{Pop}},
  \bibinfo{author}{\bibfnamefont{I.} \bibnamefont{Matei}},
  \bibinfo{author}{\bibfnamefont{\'E.} \bibnamefont{Dumur}},
  \bibinfo{author}{\bibfnamefont{A.~K.} \bibnamefont{Feofanov}},
  \bibinfo{author}{\bibfnamefont{C.} \bibnamefont{Naud}},
  \bibinfo{author}{\bibfnamefont{W.} \bibnamefont{Guichard}},
  \bibnamefont{and} \bibinfo{author}{\bibfnamefont{O.}~\bibnamefont{Buisson}},
  \bibinfo{journal}{Phys. Rev. Lett.} \textbf{\bibinfo{volume}{108}},
  \bibinfo{pages}{107001} (\bibinfo{year}{2012}).

\bibitem[{\citenamefont{Srinivasan et~al.}(2011)\citenamefont{Srinivasan,
  Hoffman, Gambetta, and Houck}}]{Srinivasan11}
\bibinfo{author}{\bibfnamefont{S.~J.} \bibnamefont{Srinivasan}},
  \bibinfo{author}{\bibfnamefont{A.~J.} \bibnamefont{Hoffman}},
  \bibinfo{author}{\bibfnamefont{J.~M.} \bibnamefont{Gambetta}},
  \bibnamefont{and} \bibinfo{author}{\bibfnamefont{A.~A.} \bibnamefont{Houck}},
  \bibinfo{journal}{Phys. Rev. Lett.} \textbf{\bibinfo{volume}{106}},
  \bibinfo{pages}{083601} (\bibinfo{year}{2011}).

\bibitem[{\citenamefont{Hoffman et~al.}(2011)\citenamefont{Hoffman, Srinivasan,
Gambetta, and Houck}}]{Hoffman2011}
\bibinfo{author}{\bibfnamefont{A.~J.}~\bibnamefont{Hoffman}},
  \bibinfo{author}{\bibfnamefont{S.~J.}~\bibnamefont{Srinivasan}},
  \bibinfo{author}{\bibfnamefont{J.~M.} \bibnamefont{Gambetta}},
  \bibnamefont{and} \bibinfo{author}{\bibfnamefont{A.~A.}~\bibnamefont{Houck}},
  \bibinfo{journal}{Phys. Rev. B} \textbf{\bibinfo{volume}{84}},
  \bibinfo{pages}{184515} (\bibinfo{year}{2011}).

\bibitem[{\citenamefont{Lecocq et~al.}(2011{\natexlab{a}})\citenamefont{Lecocq,
  Claudon, Buisson, and Milman}}]{Lecocq11squid}
\bibinfo{author}{\bibfnamefont{F.}~\bibnamefont{Lecocq}},
  \bibinfo{author}{\bibfnamefont{J.}~\bibnamefont{Claudon}},
  \bibinfo{author}{\bibfnamefont{O.}~\bibnamefont{Buisson}}, \bibnamefont{and}
  \bibinfo{author}{\bibfnamefont{P.}~\bibnamefont{Milman}},
  \bibinfo{journal}{Phys. Rev. Lett.} \textbf{\bibinfo{volume}{107}},
  \bibinfo{pages}{197002} (\bibinfo{year}{2011}{\natexlab{a}}).

\bibitem[{\citenamefont{Wallraff et~al.}(2004)\citenamefont{Wallraff, Schuster,
  Blais, Frunzio, Huang, Majer, Kumar, Girvin, and Schoelkopf}}]{Wallraff04}
\bibinfo{author}{\bibfnamefont{A.}~\bibnamefont{Wallraff}},
  \bibinfo{author}{\bibfnamefont{D.~I.} \bibnamefont{Schuster}},
  \bibinfo{author}{\bibfnamefont{A.}~\bibnamefont{Blais}},
  \bibinfo{author}{\bibfnamefont{L.}~\bibnamefont{Frunzio}},
  \bibinfo{author}{\bibfnamefont{R.-S.} \bibnamefont{Huang}},
  \bibinfo{author}{\bibfnamefont{J.}~\bibnamefont{Majer}},
  \bibinfo{author}{\bibfnamefont{S.}~\bibnamefont{Kumar}},
  \bibinfo{author}{\bibfnamefont{S.~M.} \bibnamefont{Girvin}},
  \bibnamefont{and} \bibinfo{author}{\bibfnamefont{R.~J.}
  \bibnamefont{Schoelkopf}}, \bibinfo{journal}{Nature (London)}
  \textbf{\bibinfo{volume}{431}}, \bibinfo{pages}{162} (\bibinfo{year}{2004}).

\bibitem[{\citenamefont{Plantenberg et~al.}(2007)\citenamefont{Plantenberg,
  de~Groot, Harmans, and Mooij}}]{Plantenberg07}
\bibinfo{author}{\bibfnamefont{J.~H.} \bibnamefont{Plantenberg}},
  \bibinfo{author}{\bibfnamefont{P.~C.} \bibnamefont{de~Groot}},
  \bibinfo{author}{\bibfnamefont{C.~J.
P.~M.} \bibnamefont{Harmans}},
  \bibnamefont{and} \bibinfo{author}{\bibfnamefont{J.~E.} \bibnamefont{Mooij}},
  \bibinfo{journal}{Nature} \textbf{\bibinfo{volume}{447}},
  \bibinfo{pages}{836} (\bibinfo{year}{2007}).

\bibitem[{\citenamefont{Lecocq et~al.}(2011{\natexlab{b}})\citenamefont{Lecocq,
  Pop, Peng, Matei, Crozes, Fournier, Naud, Guichard, and
  Buisson}}]{Lecocq11ebeam}
\bibinfo{author}{\bibfnamefont{F.}~\bibnamefont{Lecocq}},
  \bibinfo{author}{\bibfnamefont{I.~M.} \bibnamefont{Pop}},
  \bibinfo{author}{\bibfnamefont{Z.}~\bibnamefont{Peng}},
  \bibinfo{author}{\bibfnamefont{I.}~\bibnamefont{Matei}},
  \bibinfo{author}{\bibfnamefont{T.}~\bibnamefont{Crozes}},
  \bibinfo{author}{\bibfnamefont{T.}~\bibnamefont{Fournier}},
  \bibinfo{author}{\bibfnamefont{C.}~\bibnamefont{Naud}},
  \bibinfo{author}{\bibfnamefont{W.}~\bibnamefont{Guichard}}, \bibnamefont{and}
  \bibinfo{author}{\bibfnamefont{O.}~\bibnamefont{Buisson}},
  \bibinfo{journal}{Nanotechnology} \textbf{\bibinfo{volume}{22}},
  \bibinfo{pages}{315302} (\bibinfo{year}{2011}{\natexlab{b}}).

\bibitem[{\citenamefont{Osborn et~al.}(2007)\citenamefont{Osborn, Strong,
  Sirois, and Simmonds}}]{Osborn07}
\bibinfo{author}{\bibfnamefont{K.}~\bibnamefont{Osborn}},
  \bibinfo{author}{\bibfnamefont{J.~A.} \bibnamefont{Strong}},
  \bibinfo{author}{\bibfnamefont{A.}~\bibnamefont{Sirois}}, \bibnamefont{and}
  \bibinfo{author}{\bibfnamefont{R.}~\bibnamefont{Simmonds}},
  \bibinfo{journal}{Applied Superconductivity, IEEE Transactions on}
  \textbf{\bibinfo{volume}{17}}, \bibinfo{pages}{166} (\bibinfo{year}{2007}).

\bibitem[{\citenamefont{Groth et~al.}(2014)\citenamefont{Groth, Wimmer,
  Akhmerov, and Waintal}}]{Groth14}
\bibinfo{author}{\bibfnamefont{C.~W.} \bibnamefont{Groth}},
  \bibinfo{author}{\bibfnamefont{M.}~\bibnamefont{Wimmer}},
  \bibinfo{author}{\bibfnamefont{A.~R.} \bibnamefont{Akhmerov}},
  \bibnamefont{and} \bibinfo{author}{\bibfnamefont{X.}~\bibnamefont{Waintal}},
  \bibinfo{journal}{New Journal of Physics} \textbf{\bibinfo{volume}{16}},
  \bibinfo{pages}{063065} (\bibinfo{year}{2014}).

\bibitem[{\citenamefont{Claudon et~al.}(2008)\citenamefont{Claudon, Zazunov,
  Hekking, and Buisson}}]{Claudon08}
\bibinfo{author}{\bibfnamefont{J.}~\bibnamefont{Claudon}},
  \bibinfo{author}{\bibfnamefont{A.}~\bibnamefont{Zazunov}},
  \bibinfo{author}{\bibfnamefont{F.~W.~J.} \bibnamefont{Hekking}},
  \bibnamefont{and} \bibinfo{author}{\bibfnamefont{O.}~\bibnamefont{Buisson}},
  \bibinfo{journal}{Phys. Rev. B} \textbf{\bibinfo{volume}{78}},
  \bibinfo{pages}{184503} (\bibinfo{year}{2008}).

\end{thebibliography}
\end{document}